\documentclass[reprint]{revtex4-1}
\usepackage{amsmath, amssymb, graphics, graphicx,fullpage, mdwlist}
\newcommand{\mathsym}[1]{{}}
\newcommand{\unicode}[1]{{}}

\begin{document}

\title{A minimum control ancilla driven quantum computation scheme with repeat-until-success style gate generation}
\author{Kerem Halil-Shah, Daniel K. L. Oi}
\affiliation{SUPA Department Of Physics, University of Strathclyde\\
107 Rottenrow, Glasgow G4 0NG, UK\\
k.halil-shah@strath.ac.uk, daniel.oi@strath.ac.uk }

\begin{abstract}

Some two qubit interactions are singly sufficient for universal quantum computation but not without the use of an ancilla. Recent schemes for universal quantum computation have focused on hybrid physical systems using ancillae. In them, the application of resources is shifted to the ancilla system. We consider which 2-qubit interactions are universal in ancilla schemes where direct connections between main register qubits are forbidden. By the use of ancilla driven operations and repeat-until-success style random gates, a single fixed symmetric gate can be universal be control of the number of repetitions alone.
\end{abstract}
\maketitle


\section{Introduction}

A finite set of operations is said to be universal for quantum computation if they can be used to approximate any arbitrary unitary operation. The study of resources for quantum computation looks at finding different universal operation sets and relating them to potential physical implementations. Some of the first proofs of universality involved a single quantum gate operation: the three qubit Deutsch gate \cite{Deutsch1989} was a generalisation of the classically universal Toffoli gate; later, gates that could be applied to only two qubits were shown to be universal by Barenco \cite{Barenco1995} and Sleator \& Weinfurter \cite{SleatorWeinfurter} by showing that they could be used to recreate the Deutsch gate by successive applications to different pairs, with a minimum of five applications \cite{YuDuanYing}. One could create single qubit or two qubit gates by using ancillary qubits and generating the subgroups $U_2 \otimes \mathbb{I}_4$ and $U_4 \otimes \mathbb{I}_2$.

DiVincenzo \cite{DiVincenzo1995} showed that it was possible to decomposed any unitary operation into two qubit operations on different pairs and that the ability to perform any arbitrary two qubit unitary was universal. Deutsch \emph{et al} \cite{DeutschBarencoEkert} and Lloyd \cite{Lloyd1995} showed that in fact almost any two qubit gate is universal for two qubit unitary gates; all those that aren't have been characterised \cite{ChildsLMO11} though those which are not universal with ancilla have not been completely characterised. A key part of using a single fixed gate is that one should be able to swap the qubits with respect to the interaction and then rely upon a lack of SWAP symmetry in the two qubit gate interaction. One would not expect to be able to create a gate that consists of two different single qubit gates on each qubit if one did not have this swapping ability and swap asymmetry.

Then it was shown by Barenco et al \cite{ElementaryGatesForQC} that any arbitrary two qubit gate could be made from a single CNOT gate (later generalised to any two qubit entangling gate \cite{QCWithAnyEntanglingGate}) and arbitrary single qubit gates. This allows one to rely upon a single fixed symmetric interaction between two qubit and then apply control through the application of single qubit gates.

\begin{figure}[h]
\includegraphics[height=0.4\columnwidth,width=\columnwidth]{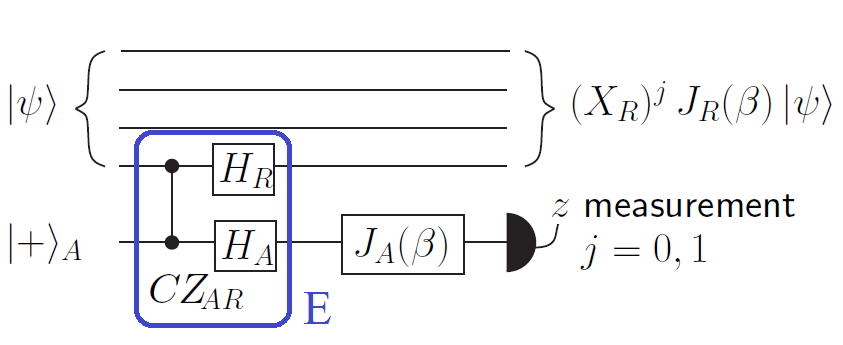}
\caption{An example of the implementation of the ancilla driven scheme from \cite{ADQC}. The interaction CZ itself is not sufficient, the inclusion of the local $H_A\otimes H_R$ gates allows the application of the gate $J(\beta)=H.e^{-i\frac{\beta}{2}\sigma_z}$ on the ancilla to result in the same gate on the register up to Pauli post-corrections. Several applications of $J(\beta)$ with different choices of $\beta$ can compose any arbitrary unitary up-to a global phase and a single Pauli post-correction \cite{ParsimoniousAndRobust}}
\label{fig:ADQCpaperSingleQubit}
\end{figure}

\begin{figure}[h]
\includegraphics[height=0.3\columnwidth,width=\columnwidth]{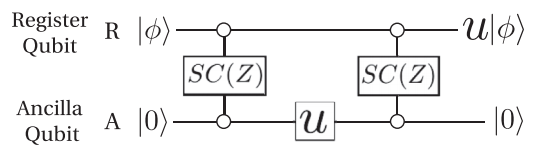}
\caption{A simple example of ancilla controlled quantum computation from \cite{ACQC}. $SC(u)$ represents the class of interactions equivalent to a SWAP gate and a Control-$U$ gate. By choice of the prepared state of the ancilla and two applications of the interaction, the scheme avoids the use of ancilla measurements, similar to quantum bus scheme without homodyne detection.}
\label{fig:ACQCpaper}
\end{figure}

 Recently a subset of schemes have arisen based around the use of ancilla systems, such as the ancilla-driven \cite{ADQC}, ancilla-control \cite{ACQC} and quantum bus (qubus) \cite{QubusProposal} scheme, where operations that match the resources of the gate circuit scheme are generated by interacting the qubits of the main register with an ancilla system then performing operations on that ancilla system. Ancilla-control quantum computation (ACQC) is characterised by swapping the register state onto the ancilla, performing local unitary gates and then swapping it back. In contrast ancilla driven quantum computation (ADQC) is marked by the evolution of the register being driven by measurements of the ancilla. The qubus aproach uses a coherent state as a mediator, using homodyne measurement to drive single qubit operations and multiple interactions with the bus between pairs to create two qubit interactions.
In each scheme, arbitrary single qubit gates can be kept to the ancilla system and the main memory register qubit do not directly interact but interact through the ancilla system (see figures \ref{fig:ADQCpaperSingleQubit},\ref{fig:ACQCpaper}). 

Different interactions are valid under different schemes, with some overlap. The ancilla control scheme can use CZ.SWAP or a weaker C$(u)$.SWAP gate or gates that are equivalent upto local unitary gates while ADQC can use a CZ or CZ.SWAP equivalent gate. Evolution can be ancilla driven with a weak interaction if one adopts a different scheme which moves from an exact deterministic universality to a stochastic approximate one \cite{KHSOi01}.

These ancilla schemes do not allow nor require capabilities assumed in previous proofs of universality of single unitaries such as swapping the qubits with respect to the interaction or being able to apply interactions between any pair of qubits. This is because they are optimised for the use of hybrid physical systems where the ancilla qubits and register qubits are of different physical implementations.  Often there is a trade off in the physical properties of systems that stable systems with long coherent times are achieved by having little accesibility and environmental interaction. On the one hand, the memory register in an ancilla scheme may be more stable because it does not have to be directly accessed nor does the design have to allow for direct interaction between the register qubits. On the other hand, these schemes is allow for the ancilla systems to be short-lived relative to the main memory register. This makes ancilla driven models apt for investigating the potential of systems that involve low control interaction with fixed local unitary actions such as in scattering spins in graphene \cite{FlyingQubitsInGraphene} and potentially other hybrid systems that involve flying and static qubits \cite{FlyingQubits}.

If the ancilla parameters can be fixed, one may benefit from the engineering of many ancillae with greater precision. This motivates a search for an interaction where an ancilla can be used in a manner consistent with the physical hybrid paradigm but we do not require unitary control of the ancilla. Instead the interaction should be sufficient on its own for universality alongside the ability to generate a very large number of ancillae. The problem is that the conditions for single qubit gates on the register and for two qubit gates on the register are often different in the ancilla schemes and are compensated for using the control over the local unitary gates. It is also hard to create any asymmetry in an interaction between different register qubits without single qubit gates because every register qubit has to interact with the ancilla system in the same way.
 
There is a solution in using the framework for ADQC with arbitrary entangling strength \cite{KHSOi01}. Ancilla driven operations and repeat-until-success gate implementations with random gate numbers have been used to improve the resource costs of quantum computer gate circuits \cite{PaetznickSvore}. They have also been considered for utilising imperfect interactions for communication over long distances \cite{KHSOi02}. We will use similar operations to provide a constructive proof of the universality of a class of symmetric gates. Unlike the previous ADQC schemes, the universality can not be considered as enacting arbitrary single qubit gates and a single two qubit entangling operation, in that case an asymmetric interaction is required. Either way, universality can be achieved without exerting any local control over the ancilla when the evolution driven by the ancilla measurement can produce two non-commuting gates. The measurement basis, interaction and ancilla preparation are all pre-specified in a single procedure, to be repeated, then the only discrimination between gates is the choice of when to stop reapplying the procedure: quantum control has been exchanged for classical control.

\section{The minimal control scheme}
Given a memory register of qubits onto which one wishes to perform a quantum gate circuit to enact a universal quantum computer, under this scheme one only needs to be able to interact each register qubit individually with multiple ancillary qubits via a single specific interaction $E= (H_A \otimes H_R).C(Z^{\frac{1}{2}})$:
\begin{equation}
C(Z^{\frac{1}{2}})=C(``\frac{\pi}{4}")=\text{diag}(1,1,e^{-i\frac{\pi}{4}},e^{-i\frac{\pi}{4}})
\end{equation}
The ancillae will be prepared and measured in the same specific state and the measurement basis that includes it: $|\text{+}i\rangle=\frac{1}{\sqrt{2}}(|0\rangle+i|1\rangle)$ and $\{|j\rangle =\frac{1}{\sqrt{2}}(|0\rangle+(-1)^j i|1\rangle)\}$. An individual ancilla interacts with the register only once or twice and only once per register qubit.

\begin{figure}[h]
\includegraphics[height=0.4\columnwidth,width=\columnwidth]{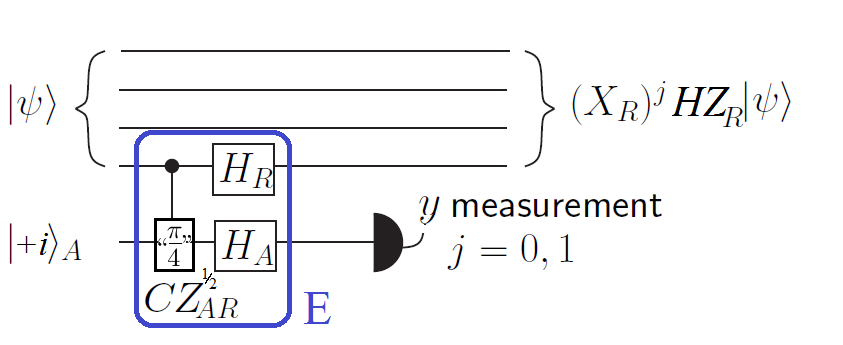}
\caption{The single step on a single qubit in the minimal control scheme; it follows the same process as ancilla driven quantum computation and the same limitation to one interaction per qubit per ancilla but with a different preparation and measurement basis.}
\label{fig:SingleQubit}
\end{figure}

Immediately before ancilla measurement, the evolution of the system can be described with the Kraus operator formalism
\begin{equation}
|\Psi\rangle\langle \Psi| \rightarrow \sum_j K_j |\Psi\rangle\langle \Psi| K^\dagger
\end{equation}
and from the Stinespring dilation theorem
\begin{equation}
K_J=\langle j| E |a\rangle . 
\end{equation} After projection of the ancilla into a specific final state $|j\rangle$ then the system has evolved $$|\Psi\rangle\langle \Psi| \rightarrow \frac{ K_j |\Psi\rangle\langle \Psi| K^\dagger}{\text{Tr}[K^\dagger_j K_j |\Psi\rangle\langle \Psi|]}.$$ Yet the pre-measurement description is independent of any choice of basis $\{|j\rangle\}$, this works because the Kraus operators follow the same rules of linear addition as the states $\{|j\rangle\}$. The final operation after a measurement result of $|m\rangle$ can be constructed from natural results of a computational basis measurement, identity or$Z^{\frac{1}{2}}\equiv \text{diag}(e^{-i\frac{\pi}{4}},e^{i\frac{\pi}{4}})$, following from the rule of linear addition:
\begin{align}
 K_{\text{+}i}&=\frac{1}{2}\left(\mathbb{I}+Z^{\frac{1}{2}}\right)=\text{cos}\left(\frac{\pi}{8}\right) Z^{\frac{1}{4}} \\
K_{\text{-}i}&=\frac{1}{2}\left(\mathbb{I}-Z^{\frac{1}{2}}\right)=i\text{sin}\left(\frac{\pi}{8}\right) Z.Z^{\frac{1}{4}}
\end{align}

Including the local Hadamard gate effects, the normalised unitary gates are $X^jHZ^{\frac{1}{4}}$. If the gate was $(H\otimes H). C(Z^{\frac{1}{n}})$, the resulting gate would be $X^jHZ^{\frac{1}{2n}}$. The two potential actions may form a universal set for single qubit gates depending on the value of $n$. In this particular case, established proofs about the universality of $\{H, Z^{\frac{1}{4}}\}$  from Boykin \emph{et al} \cite{OnUniversalAndFaultTolerant} lead to the universality of the generated pair. Two applications of the process will generate either $HZ^{\frac{1}{4}}HZ^{\frac{1}{4}}=X^{\frac{1}{4}}Z^{\frac{1}{4}}$ or $ X^{-\frac{1}{4}} Z^{\frac{1}{4}}$ up to some Pauli operator correction that will affect future pairs of operations. From Boykin \emph{et al} \cite{OnUniversalAndFaultTolerant}, it is known that the latter will produce a unitary equivalent to a rotation by an angle that is an irrational multiple of $\pi$ 
\begin{align}
U_{-+}&=e^{-i\lambda \hat{n}.\hat{\sigma}} \\
\text{cos}(\lambda \pi)&=\frac{1}{2}\left(1+\frac{1}{\sqrt{2}}\right) \\
\hat{n}&\propto\text{cot}\left(\frac{\pi}{8}\right)(\hat{z}-\hat{x})+\hat{y}
\end{align}
and then the difference in sign of the former gate only results in a change to a difference rotation axis, non-parallel to the latter, by the same irrational multiple thus together these two gates could be used to create an approximation to any single qubit unitary using a two axis decomposition. This argument relies on the ability of an irrational multiple of $\pi$ to reach any angle on a compact curve after a sufficiently large number of multiples. More generally, if the Lie algebra closure of a pair of gates covers SU(2) then these two gates form a universal set for single qubit unitary gates; the stochastically generated sequence $\prod_k U_{i(k)}$ will perform a random walk on the compact set of unitaries and be guaranteed by Poincar\'{e} recurrence to reach an approximation of any unitary eventually. The generation of the strings $\prod_k U_{i(k)}$, though probabilistic, are still consistent with the conditions of the Solovay-Kitaev theorem \cite{NielsenChuang} and so the efficiency of these approximations might be improved. So applying any single qubit unitary up to a global phase can be done through this process by applying it a large, random, number of times until the product of all the resulting actions is within a desired error bound such as the trace distance for single qubit gates, $|\langle U,U_T\rangle|^2=1-\frac{1}{2}|\text{Tr}[U^\dagger U_T]|$ where $U_T$ is the target unitary.

One can derive an upper bound on the expected time by considering the expected time to hit the gates in a universal gate set e.g. $\{H,T\}$. After decomposing a target gate into the finite gate set, the number of times the gates appear in that decomposition can be taken and multiplied by their expected times to give the expected time it would take to attempt the target gate by first waiting until we have achieved each individual gate in the decomposition, in sequence. Of course the actual expected time of the target may be less than that because there may be the possibility of achieving the target before completing the sequence. This raises an interesting question of how using different decompositions with different expected times reveals different upper bounds on a target's decomposition time.

\subsection{Two qubit gate}

When the ancilla interacts with a register qubit via a symmetric interaction $e^{i\alpha\sigma_z\otimes\sigma_z}$, the back-action on the ancilla produces a mixture of a pair of states symmetric about the initial state of the ancilla:
\begin{align}
|a\rangle\sum_j c_j |j\rangle \rightarrow& \sum_j c_j |a_j\rangle|j\rangle, \\
 |a_j\rangle=&R_{\hat{z}}(\pm 2\alpha)|a\rangle \\
K_m =\langle m|a_0\rangle\otimes |0\rangle\langle 0|+&\langle m|a_1\rangle\otimes|1\rangle\langle 1|
\end{align}
In order for the final action on the register qubit to be unitary, the ancilla measurement basis must be symmetric with respect to the back-action induced states. Similarly in the the two qubit gate process, there is a four state description:
\begin{equation*}
|a\rangle\sum_{jk} c_{jk} |j\rangle|k\rangle \rightarrow \sum_{jk} c_{jk} |a_{jk}\rangle|j\rangle|k\rangle,
\end{equation*}
in addition to which is the requirement that the planes of symmetry from each interaction with an individual qubit are perpendicular \cite{KHSOi01}. 

Thus the single qubit process needs measurement in the basis that includes the initial ancilla state while the two qubit process requires some local unitary gate actions on the ancilla that are rotations about that basis to the measurement. With the initial state of the ancilla being a vector along the $\hat{y}$ axis of the Bloch sphere, the Hadamard gate fulfils this condition by rotating states $|a_j\rangle$ in the $\hat{x}-\hat{y}$ plane into the $\hat{z}-\hat{y}$ plane while preserving symmetry around the $\hat{y}$ axis. So with the same gate, the same process can be used to implement a two qubit entangling gate between two different register qubits by allowing the qubit to interact with each qubit while being prepared and measured in the same basis.

\begin{figure}[h]
\includegraphics[height=0.4\columnwidth,width=\columnwidth]{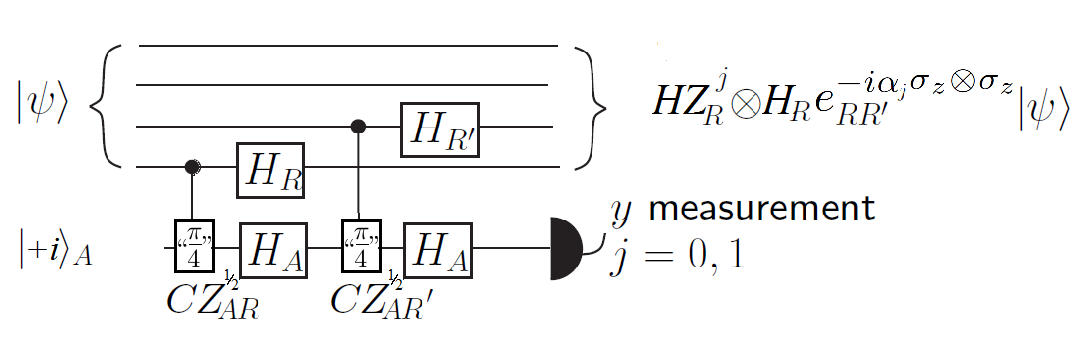}
\caption{The schematic for performing a two qubit gate under minimal control. An identical interaction with a second register qubit is turned on but all other processes remain the same.}
\label{fig:TwoQubit}
\end{figure}

This time, the resulting Kraus operators will be a two qubit gate that can be written as a symmetric interaction $e^{-i\alpha \sigma_z\otimes\sigma_z}$ with the same local gates. Due to the ability to perform single qubit gates under the process detailed in the previous section, one can assume that the local gates can be undone and the description can be continued with only the symmetric non-local part of the decomposition. The $+i,-i$ results will correspond to the generation of the unitary operations $e^{i(\phi+\frac{\pi}{4})\sigma_z\otimes\sigma_z}$ and $e^{-i\frac{\pi}{4}\sigma_z\otimes\sigma_z}$ where $\text{tan}(\phi)=\frac{-1}{\text{cos}(\frac{\pi}{4})}=-\sqrt{2}$. As can be seen from the decomposition, interactions of this class can map to a Control-Unitary gate $CR_{\hat{z}}(4\alpha)$ and accordingly form a compact group that can be represented by a closed 1-d curve. Randomly generating gates in this class randomly walks over the curve. In this case, $\phi$ is an irrational multiple of $\pi$, provable from the theorem that $\frac{1}{\pi} \text{arccos}(\frac{1}{\sqrt{n}})$ is irrational for odd $n\geq 3$ \cite{TheBookP31}. Again, a target unitary can be chosen such as $e^{i\frac{\pi}{4}\sigma_z\otimes\sigma_z}$ ,equivalent to CZ, and the process repeated until the product of repeated generations approaches within a distance of the target.

\begin{figure}[h]
\includegraphics[height=0.44\columnwidth,width=\columnwidth]{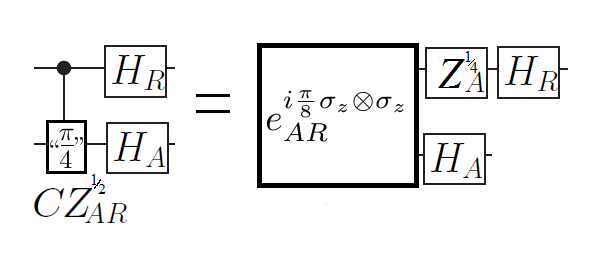}
\caption{The Cartan decomposition of the gate in the minimal control scheme reveals that it is asymmetric in its local gate actions.}
\label{fig:CartanDecomposition}
\end{figure}

The interaction can be, for illustrative purposes, re-written as a symmetric non-local action $e^{-i\frac{\pi}{8}\sigma_z\otimes\sigma_z}$ and an asymmetric pair of local actions on each subsystem. This asymmetry is key to the result. When the gate has been prepared in a Pauli operator eigenstate, this will mean that the local gate actions on the ancilla after the symmetric interaction will be Clifford group single qubit gates. However, universality for the single qubit gates requires that there is a non-Clifford gate acting on the register qubit side, making an asymmetry in the gate construction. It also means that the process includes a universal gate set for direct register manipulation. What we will demonstrate in the following section is that it is also possible to use a completely symmetric interaction with Hadamard local gates by using additional register qubits as an ancillary system.

\section{Converting symmetry gates into asymmetric gates on the register}

Given a symmetric ancilla-register gate, we can not use the universality of a fixed two qubit entangling gate with arbitrary single qubit gates to demonstrate the universality of a classical control ancilla scheme as with other schemes. However it is still possible to achieve universality by constructing an asymmetric gate on the register.

To fulfill the condition for the gate to operate on both single and pair register qubits, we know it must be of the form  $(U_L\otimes U_L)e^{i\alpha \sigma_z\otimes\sigma_z}$  where $U_L$ is a single qubit Clifford group operation; let us consider $U_L=H$. By ancilla driven operations, this can provide the finite gate set generated by $\{H,Z\}$ stochastically. With this gate set, we can correct the local unitary effects on the register qubits. Performing a random walk in this finite graph will be much faster than over the 3-d parameters of $SU(2)$. This secures the ability to generate a random walk through the 1-d curve representing the interaction $e^{i\beta \sigma_z \otimes \sigma_z}$.

One can perform the random walk carried out by the gates $e^{-i(\phi+\frac{\pi}{4})\sigma_z\otimes\sigma_z}$ and $e^{-i\frac{\pi}{4}\sigma_z\otimes\sigma_z}$ ($\text{Tan}(\phi)=\frac{-1}{\text{cos}(2\alpha)}$) until it has reached an appropriate target $\beta$. The exact stopping condition will depend on $\alpha$ and $\phi(\alpha)$; if the value of $\phi$ is an irrational multiple of $\pi$ then it can be used to make an approximation but if it is not then it can exactly generate a finite group of fractions of $\pi$ e.g. $\beta=\frac{\pi}{8}$ which is particularly of interest since it is locally equivalent to the gate diag$(1,1,1,i)$ (equivalent to but not exactly $C(Z^{\frac{1}{2}})$) which is universal when combined with $H$ \cite{OnUniversalAndFaultTolerant}. We can in fact construct an asymmetric gate $(H\otimes \mathbb{I}).e^{-i\frac{\pi}{8}\sigma_z\otimes\sigma_z}$ and utilise the universality over two qubits of asymmetric gates of non-zero trace.

\section{Summary}

Almost all two qubit gates are universal when we allow exchange of qubits with respect to the gate, interactions between any pair of qubits and the use of ancilla qubits. However when ancilla schemes are constrained by the use of a physical hybrid, it was necessary to employ a fixed gate and local unitary gate control of the ancilla. Ancilla driven operations and repeat-until-success gate implementations with random gate numbers have been used to improve the resource costs of quantum computer gate circuits \cite{PaetznickSvore}. They have also been considered for utilising imperfect interactions for communication over long distances \cite{KHSOi02}. We have demonstrated how an ancilla driven scheme for universal quantum computation can use a single fixed gate $H\otimes H.e^{i\alpha \sigma_z \otimes \sigma_z}$ if one adopts repeat-until-success style random gate generation. 

The authors would like to thank Joseph Fitzsimmons, Aminesh Datta, Elham Kashefi and Erika Anderson for their advice in the preparation of this manuscript. This work was funded by the Engineering and Physical Sciences Research Council (EPSRC), partly supported by Quantum Information Scotland (QUISCO).

\bibliographystyle{unsrt}
\bibliography{bibliography}

\end{document}